\documentclass{aastex61}

\usepackage{natbib}
\usepackage{apjfonts}
\usepackage{amsmath}

\newcommand{\lae}{\mathrel{\raise .4ex\hbox{\rlap{$<$}\lower 1.2ex\hbox{$\sim$}}}}
\newcommand{\gae}{\mathrel{\raise .4ex\hbox{\rlap{$>$}\lower 1.2ex\hbox{$\sim$}}}}

\def\arcsec{\hbox{$^{\prime\prime}$}}
\def\deg{\hbox{$^\circ$}}

\newcommand{\ixpe}{\mbox{\em IXPE\/}}

\shorttitle{Further Development of X-ray Polarimetry Analysis}
\shortauthors{Marshall}

\begin{document}

\submitjournal{Astrophysical Journal}
\received{September 8, 2023}
\accepted{October 30, 2023}

\title{Further Development of Event-Based Analysis of X-ray Polarization Data}

\correspondingauthor{Herman L.\ Marshall}
\email{hermanm@space.mit.edu}

\author{Herman L. Marshall}
\affiliation{Kavli Institute for Astrophysics and Space Research,
 Massachusetts Institute of Technology, 77 Massachusetts Ave.,
 Cambridge, MA 02139, USA}

\begin{abstract}
An event-based maximum likelihood method for handling X-ray polarimetry data is extended to include the effects of background and nonuniform sampling of the possible position angle space.
While nonuniform sampling in position angle space generally introduces cross terms in the uncertainties of polarization parameters that could create degeneracies, there are interesting cases that engender no bias or parameter covariance.
When including background in Poisson-based likelihood formulation, the formula for the minimum detectable polarization (MDP) has nearly the same form as for the case of Gaussian statistics derived by \cite{2012SPIE.8443E..4NE} in the limiting case of an unpolarized signal.
{ A polarized background is also considered, which demonstrably increases uncertainties in source polarization measurements.}
In addition, a Kolmogorov-style test of the event position angle distribution is proposed that can provide an unbinned test of models where the polarization angle in Stokes space depends on event characteristics such as time or energy.
\end{abstract}


\section{Introduction}

The goal of this paper is to extend the maximum likelihood formulation developed earlier for analysis of unbinned X-ray polarimetry data \citep{marshall20} to circumstances that were not considered there.  The method was developed specifically for application to data from the Imaging X-ray Polarization Explorer \cite[IXPE,][]{weisskopf2022} but can be applied generally to instruments that yield events with associated polarization information, such as a soft X-ray polarimeter \citep{redsoxjatis} that is now in development, or instruments that must be rotated to obtain polarization information.  In the case of IXPE, there is an angle $\psi$ associated with every event based on the track produced by the photoelectron ejected by the incident X-ray.  For the soft X-ray polarimeter, each event is associated with a ``channel'' according to the position 
angle of its Bragg reflector relative to the sky.

{ By design, the gas pixel detectors on \ixpe\ \citep{2023AJ....165..186R}
and PolarLight \citep{2019ExA....47..225F} have uniform sensitivity with $\psi$. This
is not generally true for systems based on
Bragg reflection \citep[e.g. OSO-8,][]{1976ApJ...208L.125W}, Thomson scattering \citep[e.g. POLIX on XPoSat,][]{2022cosp...44.1853P}, or Compton scattering  \citep[e.g. X Calibur,][]{2014JAI.....340008B}.  Such instruments
usually require rotation to obtain uniform azimuthal
exposure.  See the review of instruments based on Compton scattering by
\citet{DelMonte2022}.
Thus, in section \ref{sec:asymmetry}, exposure nonuniformities are examined and characterized by two observation-based
parameters that can be used to determine the impact of such asymmetries.

Every instrument has a background signal, so in section~\ref{sec:modelbg}, a background term is added to the unbinned likelihood model.
The basic case of an unpolarized signal is covered in section~\ref{sec:unpol_modelbg} and augmented to include
the impact of unpolarized background in section~\ref{sec:pol_modelbg}.

Given a model with its best fit parameters, it is necessary to test it.
A Kolmogorov test of the counts with time or energy would not be sensitive to the polarization model.  Previous tests
of polarization models generally examined only the significances of the estimates of the polarization fraction
for a full observation \citep[e.g.][]{2022Natur.611..677L} or perhaps when binned by energy or pulse phase \citep[e.g.][]{2022Sci...378..646T}.
In section \ref{sec:modeltest}, a new test is proposed that is specifically designed to be sensitive to whether the distribution
of the event $\psi$ values matches the model.  This sort of test can be used to examine the validity of a pulsar rotating vector
model, such as fit by the unbinned method developed by \citet{2023MNRAS.519.5902G}.  This test method can also be useful
in cases where the electric vector position angle (EVPA) rotates with time as in two observations of the BL Lac object Mk 421 \citep{mk421rot} in order to test whether the rotation occurs at a uniform rate without binning EVPA measurements in time.}

A short review of the maximum likelihood formalism is in order, following \citet{marshall20} and \citet{2021AJ....162..134M}.
For this analysis, consider a simple case of a fixed energy band over which the polarization is constant so that the data consist of counts in $\psi$ space.
At energy $E$, the modulation factor of the instrument is $\mu_E$, the instrument effective area is $A_E$, and the intrinsic source photon flux is $f_E$ based on the spectral model of the source.  Both $\mu_E$ and $A_E$ are assumed to be known {\it a priori}.  The event density in a differential energy-phase element $dE d\psi$ about $(E, \psi)$ is

\begin{equation}
\lambda(E, \psi) = \frac{1}{2 \pi}[ 1 +  \mu_E  (q \cos 2\psi + u \sin 2\psi) ] f_E A_E T
\end{equation}

\noindent
where $T$ is the exposure time and the (normalized) Stokes parameters are $q \equiv Q/I$ and $u \equiv U/I$ for Stokes fluxes $I$, $Q$, and $U$.  (Circular polarization, $V$, is ignored here, as there is currently no practical way to measure it in the X-ray band.)

Assuming that there are $N$ events, with energies and instrument angles $(E_i, \psi_i)$, then
the log-likelihood for a Poisson probability distribution of events, $S = -2 \ln L$, is

\begin{eqnarray}
S & = & -2 \sum_i^N \ln \lambda(E_i, \psi_i) + \frac{T}{\pi} \int f_E A_E dE  \int_0^{2\pi} [ 1 +  \mu_E  (q \cos 2\psi + u \sin 2\psi) ]  d\psi \\
 & = & -2 \sum_i^N \ln f_i -2 \sum_i^N  \ln( 1 +  q \mu_i \cos 2\psi_i + u \mu_i \sin 2\psi_i) + 2 T \int f_E A_E dE
\label{eq:case1}
\end{eqnarray}

\noindent
where $f_i \equiv f(E_i)$ and $\mu_i \equiv \mu(E_i)$, after dropping terms independent of $q$, $u$, and $f$.
In this case, the log-likelihood
for the polarization parameters alone (such as when the polarization is independent of $E$) is relatively simple:
\begin{equation}
S(q,u) = -2 \sum_i^N  \ln( 1 +  q \mu_i \cos 2\psi_i + u \mu_i \sin 2\psi_i) =
-2 \sum_i^N \ln(1 + q c_i + u s_i)
\end{equation}

\noindent
where $c_i = \mu_i \cos 2\psi_i$ and $s_i = \mu_i \sin 2\psi_i$. {  For a weakly polarized source,
the best estimates of $q$ and $u$ are well approximated as
$\sum_i c_i / \sum_i c_i^2$ and $\sum_i s_i / \sum_i s_i^2$, respectively. }
  See \cite{marshall20} for details.

\section{Nonuniform Exposure}
\label{sec:asymmetry}

Now, consider the case of a nonuniform exposure in an observation of an unvarying source.
The exposure function, $w(\psi)$ with units of radians$^{-1}$, can be defined as the fraction of the exposure
spent with sensitivity to phase angle $\psi$.  If the total
exposure is $T$, then the exposure function can be normalized such that it integrates to unity for $0 \leq \psi < 2 \pi$.
In this case, the event density is

\begin{equation}
\lambda(E, \psi) = [ 1 +  \mu_E  (q \cos 2\psi + u \sin 2\psi) ] f_E A_E T dE w(\psi) d\psi
\end{equation}

\noindent
and the log-likelihood for a Poisson probability distribution of events, $S = -2 \ln L$, is

\begin{eqnarray}
S & ~=~ & -2 \sum_i \ln \lambda(E_i, \psi_i) + 2 T \int f_E A_E dE  \int_0^{2\pi} [ 1 +  \mu_E  (q \cos 2\psi + u \sin 2\psi) ]  w(\psi) d\psi
\end{eqnarray}

\noindent
To simplify some results, now assume that the spectrum has a spectral shape with uninteresting spectral shape parameters
$\xi$ that are not related to the polarization so that $f_E = f_0 \eta(E;\xi)$ and define
$K = T \int \eta(E;\xi) A_E dE$ and $K_{\mu} = T \int \eta(E;\xi) A_E \mu_E dE$ as
conversion constants (from flux units to counts or modulated counts), giving

\begin{equation}
\begin{aligned}
S(f_0, q, u) = & -2 N \ln f_0 -2 \sum_i  \ln( 1 +  q \mu_i \cos 2\psi_i + u \mu_i \sin 2\psi_i) \\
 & + 2 K f_0 + 2 K_{\mu} f_0 q \int_0^{2\pi} w(\psi) \cos 2\psi d\psi +
 2 K_{\mu} f_0 u \int_0^{2\pi} w(\psi) \sin 2\psi d\psi
\label{eq:exposure}
\end{aligned}
\end{equation}

\noindent
(dropping terms independent of $f_0$, $q$, or $u$).
Note that when $\mu$ is independent of $E$, $K_{\mu} = \mu K$.

Redefining the weights with trigonometric factors, we can simplify Eq.~\ref{eq:exposure}:

\begin{equation}
S(f_0, q, u)  ~=~ -2 N \ln f_0 - 2 \sum_i  \ln( 1 +  q c_i + u s_i) +
 2 K f_0 + 2 K_{\mu} f_0 A q +  2 K_{\mu} f_0 B u
\label{eq:exposure2}
\end{equation}

\noindent
where $\alpha(\psi) \equiv w(\psi) \cos 2\psi$ and $\beta(\psi) \equiv w(\psi) \sin 2\psi$,
and the integrals
of $\alpha$ and $\beta$ over $\psi$ are $A$ and $B$, respectively.  The quantities $A$ and
$B$ are unitless, with absolute values less than or of order unity.
Note that $f_0$ is covariant with $u$ and $q$ via the exposure weighting
terms $A$ and $B$.  These quantities are both zero when $w(\psi)$ is constant over $[0,\pi]$
or $[0,2\pi]$ but either or both can be nonzero otherwise. 

The best estimate of $f_0$
is readily determined by setting the setting $\partial S / \partial f_0$ to zero and
solving for $f_0$, giving

\begin{equation}
    \hat{f_0} = \frac{N}{K + K_\mu (Aq + Bu)}   ~~~.
\label{eq:gen0}
\end{equation}

\noindent
When $A$ and $B$ are zero or the polarization, $\Pi \equiv (q^2 + u^2)^{1/2}$
is zero, then $f_0$ is just $N/K$, as expected.  Setting
$\partial S/\partial u = 0$ and $\partial S/\partial q = 0$
to find the best estimates of $q$ and $u$ gives 

\begin{eqnarray}
\label{eq:gen1}
A K_\mu \hat{f_0}  = \sum_i \frac{c_i}{1 + \hat{q}c_i + \hat{u}s_i}  = \sum_i W_i c_i\\
B K_\mu \hat{f_0}  = \sum_i \frac{s_i}{1 + \hat{q}c_i + \hat{u}s_i}
= \sum_i W_i s_i
\label{eq:gen2}
\end{eqnarray}

\noindent
where $W_i \equiv (1 + \hat{q} c_i + \hat{u} s_i)^{-1}$.
As before, these two equations apply under quite general
circumstances but require numerical solution.
However, as in \citet{marshall20}, for $\hat{q} \ll 1$ and $\hat{u} \ll 1$, a simple approximate solution may be found, noting that $A$ and $B$ are generally of order unity, so

\begin{eqnarray}
\hat{q} \approx \frac{\sum_i c_i - A N K_{\mu}/K}{\sum_i c_i^2} ~~~\\
\hat{u}  \approx \frac{\sum_i s_i - B N K_{\mu}/K}{\sum_i s_i^2} ~~.
\label{eq:approxqu}
\end{eqnarray}

At this point, the uncertainties
in $q$ and $u$ can be derived.  All second derivatives of Eq.~\ref{eq:exposure2} are nonzero:

\begin{eqnarray}
\frac{\partial^2 S}{\partial f_0^2} & = & \frac{2 N}{f_0^2} \\
\frac{\partial^2 S}{\partial f_0 \partial q} & = & 2 K_\mu A \\
\frac{\partial^2 S}{\partial f_0 \partial u} & = & 2 K_\mu B \\
\frac{\partial^2 S}{\partial q^2} & = & \sum_i W_i^2 c_i^2 \approx \sum_i c_i^2 \\
\frac{\partial^2 S}{\partial u^2} & = & \sum_i W_i^2 s_i^2 \approx \sum_i s_i^2 \\ 
\frac{\partial^2 S}{\partial q \partial u} & = & \sum_i W_i^2 c_i s_i \approx \sum_i c_i s_i
\end{eqnarray}

\noindent
where, again, the approximations hold for $\hat{q} \ll 1$ and $\hat{u} \ll 1$.

We are most interested in the uncertainty in the polarization, $\Pi$.  We can make the
coordinate transformation from $(q,u)$ to $(\Pi,\varphi)$, where $\varphi = \frac{1}{2} \tan^{-1} (u/q)$ and
determine $S(\hat{f_0}, \Pi, \varphi)$:

\begin{equation}
S(\hat{f_0}, \Pi, \varphi)  ~=~ 2 N \ln [K + K_{\mu} \Pi (A \cos 2\varphi + B \sin 2 \varphi )]
    - 2 \sum_i  \ln [ 1 +  \Pi \mu_i  \cos ( 2\psi_i - 2\varphi ) ]
\label{eq:like_pphi}
\end{equation}

\noindent
for which the second derivative with respect to $\Pi$ is

\begin{equation}
\frac{\partial^2 S}{\partial \Pi^2}  ~=~ \frac{-2 N K_{\mu}^2 (A \cos 2\varphi + B \sin 2 \varphi )^2}{[K + K_{\mu} \Pi (A \cos 2\varphi + B \sin 2 \varphi )]^2}
    + 2 \sum_i  \frac{\mu_i^2 \cos^2(2\psi_i - 2\varphi)}{ [1 +  \Pi \mu_i  \cos ( 2\psi_i - 2\varphi ) ]^2}
\label{eq:partials2}
\end{equation}

\noindent
with a limit as $\Pi \longrightarrow 0$ and $A^2 + B^2 \ll 1$ giving

\begin{equation}
\frac{1}{\sigma_\Pi^2} \approx \sum_i  \mu_i^2 \cos^2(2\psi_i - 2\varphi)
- N K_{\mu}^2 (A \cos 2\varphi + B \sin 2 \varphi )^2/K^2
\label{eq:uncert}
\end{equation}

\noindent
The first term on the right hand side is the ``normal'', expected term that depends on
the modulation factor and the cosines of the phase angles.  The second term, however, is of great concern because it is negative definite, causing the uncertainty in $\Pi$ to
increase arbitrarily, and because it depends on the true but unknown phase.
If either $A$ and $B$ are nonzero, then the uncertainty in $\Pi$ depends upon this phase in
a way that can render statistical uncertainties difficult to compute and irregular.
Thus, an important
characteristic of a good polarimeter is designing it so that $A$ and $B$ are as close to zero as possible.
{ As stated in the introduction}, the gas pixel detectors on \ixpe\ \citep{2023AJ....165..186R}
have uniform sensitivity to phase angle for the entire exposure, so $A = B = 0$.
{ The case of a set of Bragg reflectors is worth examining.  A single reflector has an ideal angular response that is a delta function in $\psi$: $w(\psi) = \delta(\psi-\psi_0)$.  If there are $n_B$ reflectors, then $w(\psi) = 1/n_B \sum_i^{n_B} \delta(\psi-\psi_i)$.  It can be shown that when $\psi_i = \psi_0 + \pi i/n_B$, then $A$ and $B$ are identically zero for arbitrary $\psi_0$ when $n_B > 2$ and the solution to Eqs.~\ref{eq:gen0} to \ref{eq:gen2} is not degenerate}.\footnote{ For $n_B = 2$, $A = B = 0$ also, but then the system of equations becomes degenerate and no unique solution is possible.  For example, Eq.~\ref{eq:gen2} is $0=0$ for $\psi_0 = 0$.}
For the broad-band soft X-ray polarimeter with 3 Bragg reflectors at 120\deg\ to each other \citep{redsoxjatis}, $A = B = 0$ if all three channels are operated for the same time period.

\section{Adding a Background Term}

\label{sec:modelbg}

{ There are two cases to consider.  The easier case is when the background is unpolarized.  This case helps set the stage for the case
of polarized background, which is important for situations such as when measuring a pulsar inside a pulsar wind nebula or a source in the
wings of a brighter, polarized source.}

{ Regardless of whether the background is polarized,} a background region of solid angle $\Omega$ is chosen that is source free and the
source region covers a solid angle $\zeta \Omega$ that is presumed to have
the same background { characteristics}.
There are $N$ events in the source region { labeled with index $i$} and
$N_B$ events in the background region { labeled with index $j$}.
This case is similar to that considered by \citet{2012SPIE.8443E..4NE} for
the case of Gaussian counting statistics.
To compare to their analysis more directly, we expect
$C_B \equiv \zeta N_B$
counts in the source region to be due to background, giving
$N-C_B \equiv C_S$ {\em net} counts in the source region.
{ In this analysis,} the exposure is uniform over $\psi$.

\subsection{Unpolarized Background}
\label{sec:unpol_modelbg}

{ If the background is unpolarized, the event density is relatively simple:}

\begin{equation}
\lambda_S(\psi) = \frac{1}{2 \pi} \{ N_0 [ 1 +  \mu  (q \cos 2\psi + u \sin 2\psi) ] + \zeta B\}
\label{eq:lambda_bg}
\end{equation}

\noindent
for the source region and $\lambda_B(\psi) = \frac{B}{2 \pi}$ for the background region.
Here, the notation is simplified by defining $N_0 = f_0 T \int \eta(E;\xi) A_E dE$, which is just the expected number of counts from the source under some spectral model $f_0 \eta(E;\xi)$.
Then, the log-likelihood for a Poisson probability distribution of source and background events, $S = -2 \ln L$, is

\begin{eqnarray}
S & ~=~ & -2 \sum_{i=1}^{N} \ln \lambda_S(\psi_i) + \frac{1}{2 \pi} \int_0^{2\pi} [ N_0 (1 +  \mu q \cos 2\psi + \mu u \sin 2\psi) + \zeta B] d\psi - 2 \sum_{j=1}^{N_B} \ln B + 2 B\\
 & ~=~ & -2 \sum_{i=1}^{N} \ln[ N_0 (1 + q c_i + u s_i) +\zeta B] + 2 N_0 + 2 B (1+\zeta) - 2 N_B \ln B
\label{eq:bg}
\end{eqnarray}

\noindent
(dropping terms independent of $B$, $N_0$, $q$, or $u$).
Setting partial derivatives to zero gives

\begin{eqnarray}
   \label{eq:n0hat}
    \hat{N_0} & ~=~ & \sum_{i=1}^{N} \frac{1+\hat{q} c_i + \hat{u} s_i }{1+\hat{q} c_i + \hat{u} s_i+\frac{\zeta \hat{B}}{\hat{N_0}} } = \sum w_i + \hat{q} \sum w_i c_i + \hat{u} \sum w_i s_i = \sum w_i \\
    \label{eq:bcb}
    \hat{B} & ~=~ & \frac{N_B}{1 + \zeta(1-\frac{\sum w_i}{\hat{N_0}})} = N_B\\
      \label{eq:wc}
   0 & ~=~ & \sum w_i c_i \\
    0 & ~=~ & \sum w_i s_i
    \label{eq:ws}
\end{eqnarray}
for $N_0 \ne 0$ and defining $w_i = [ 1+\hat{q} c_i + \hat{u} s_i +\zeta \hat{B} / \hat{N_0}]^{-1}$.
Eqs.~\ref{eq:wc} and \ref{eq:ws} have been used to simplify Eq.~\ref{eq:n0hat} and
Eq.~\ref{eq:n0hat} is used to simplify Eq.~\ref{eq:bcb}.
Substituting
$N_B$ for $B$ in Eq.~\ref{eq:n0hat} and transforming from $(q,u)$ to $(\Pi,\varphi)$ gives
\begin{equation}
\hat{N_0} = \sum_{i=1}^{N} [1 + \hat{\Pi} \mu_i \cos (2 \psi_i-2 \hat{\varphi}) + \frac{\zeta N_B}{\hat{N_0}} ]^{-1}  ,
\label{eq:n0}
\end{equation}
which can be solved for $\hat{N_0}$ for trial values of $\hat{\Pi}$ and $\hat{\varphi}$ to make minimizing $S$ simpler by substituting $\hat{N_0}$ and $\hat{B} = N_B$ into Eq.~\ref{eq:bg}.  As $\hat{\Pi} \longrightarrow 0$
$\hat{N_0} \longrightarrow N - \zeta N_B = C_S$, as expected, providing a good starting point for estimating $\hat{N_0}$.

The minimum detectable polarization (MDP) for this case can be estimated by computing the uncertainty in $\Pi$, $\sigma_{\Pi}$, by

\begin{equation}
    \frac{2}{\sigma_{\Pi}^2} \approx  \frac{\partial^2 S}{\partial {\Pi}^2} =
    2 \sum_{i=1}^{N} w_i^2 \mu_i^2 \cos^2 (2 \psi_i-2 \varphi)
   \label{eq:ps_ppi}
\end{equation}

\noindent
Then, as $\hat{{\Pi}} \longrightarrow 0$, $w_i \longrightarrow [1 + \zeta N_B/\hat{N_0}]^{-1}$, so

\begin{equation}
    \sigma_{\Pi} \longrightarrow
    \frac{1 + \zeta N_B/\hat{N_0}}{ [\sum_{i=1}^{N} \mu_i^2 \cos^2 (2 \psi_i-2 \varphi)]^{1/2}} = 
        \frac{\sqrt{2} (1 + \zeta N_B/\hat{N_0})}{ [N \langle \mu_i^2\rangle]^{1/2}}
    = \frac{ \sqrt{2 N} }{(N - \zeta N_B) \sqrt{\langle \mu_i^2\rangle}} = \frac{ \sqrt{2 ( C_S + C_B )} }{C_S \sqrt{\langle\mu_i^2\rangle}} 
\label{eq:sigmap}
\end{equation}

\noindent
where the first step follows as $\mu_i$ and $\psi_i$ are uncorrelated and the
second step follows from the asymptotic value of $\hat{N_0}$.  Finally, the MDP at 99\% confidence is

\begin{equation}
    {\rm MDP}_{99} = 3.03 \sigma_{\Pi} = \frac{ 4.29 \sqrt{C_S + C_B} }{C_S \sqrt{\langle\mu_i^2\rangle}}  ~~,
    \label{eq:mdp_bg0}
\end{equation}
just as found by \cite{2012SPIE.8443E..4NE} for Gaussian statistics
with the exception of the substitution of the rms of $\mu_i$ for $\mu$.

{
\subsection{Polarized Background}
\label{sec:pol_modelbg}
It is more likely that the X-ray background is partially polarized as it often contains some fraction of the source as well (due to the extent of the
telescope's point spread function).
The background is assumed to be primarily due to photons, essentially indistinguishable from source events, susceptible to the same
modulation factor as source events are.
If the background is polarized, the event density has added terms giving the normalized $u$ and $q$ of the background,
denoted by $q_b$ and $u_b$:

\begin{eqnarray}
\label{eq:lambda_polbg1}
\lambda_S(\psi) & = & \frac{1}{2 \pi} \{ N_0 [ 1 +  \mu  (q \cos 2\psi + u \sin 2\psi) ] + \zeta B [ 1 +  \mu  (q_b \cos 2\psi + u_b \sin 2\psi) ] \} \\
\lambda_B(\psi) & = & \frac{B}{2 \pi} \{ 1 +  \mu  (q_b \cos 2\psi + u_b \sin 2\psi) ] \}
\label{eq:lambda_polbg2}
\end{eqnarray}

\noindent
for the source and background regions, respectively.
Then,

\begin{eqnarray}
S & ~=~ & -2 \sum_{i=1}^{N} \ln \lambda_S(\psi_i) + \frac{1}{2 \pi} \int_0^{2\pi} \lambda_S(\psi_i) d\psi
  -2 \sum_{j=1}^{N_B} \ln \lambda_B(\psi_j) + \frac{1}{2 \pi} \int_0^{2\pi} \lambda_B(\psi) d\psi\\
 & ~=~ & -2 \sum_{i=1}^{N} \ln[ N_0 (1 + q c_i + u s_i) +\zeta B(1 + q_b c_i + u_b s_i)] + 2 N_0 + 2 B (1+\zeta) - 2 N_B \ln B -2 \sum_{j=1}^{N_B} \ln (1 + q_b c_j + u_b s_j)
\label{eq:s_polbg}
\end{eqnarray}

\noindent
(dropping terms independent of $B$, $N_0$, $q$, $u$, $q_b$, or $u_b$) and again defining $c_i = \mu_i \cos 2\psi_i$ and $s_i = \mu_i \sin 2\psi_i$.
Setting partial derivatives to zero gives

\begin{eqnarray}
   \label{eq:pol_n0hat}
    \hat{N_0} & ~=~ & \sum_{i=1}^{N} \frac{1+\hat{q} c_i + \hat{u} s_i }{1+\hat{q} c_i + \hat{u} s_i+\frac{\zeta \hat{B}}{\hat{N_0}} (1+\hat{q_b} c_i +\hat{u_b} s_i) } = \sum_i^N W_i \\
    \label{eq:pol_bcb}
    \hat{B} & ~=~ & N_B\\
      \label{eq:pol_wc}
   0 & ~=~ & \sum W_i c_i \\
      \label{eq:pol_ws}
    0 & ~=~ & \sum W_i s_i \\
     \label{eq:pol_wc_bg}
   0 & ~=~ & \sum V_j c_j \\
    \label{eq:pol_vc}
    0 & ~=~ & \sum V_j s_j ~~~,
    \label{eq:pol_vs}
\end{eqnarray}
defining $W_i = [ 1+\hat{q} c_i + \hat{u} s_i +\zeta \hat{B} (1+\hat{q_b} c_i +\hat{u_b} s_i)/ \hat{N_0}]^{-1}$ and now $V_j = [1+\hat{q_b} c_j + \hat{u_b} s_j]^{-1}$.
As before, Eqs.~\ref{eq:pol_wc}, ~\ref{eq:pol_ws}, and \ref{eq:pol_n0hat} have been used to derive Eq.~\ref{eq:pol_bcb}.
Eqs.~\ref{eq:pol_vc} and \ref{eq:pol_vs} can be solved for $\hat{q_b}$ and $\hat{u_b}$ as in \citet{marshall20}, giving
\begin{eqnarray}
\hat{q_b} \approx \sum_i c_i / \sum_i c_i^2\\
\hat{u_b} \approx \sum_i s_i / \sum_i s_i^2
\end{eqnarray}
when the background is weakly polarized.  Not surprisingly, the optimal Stokes parameters for the background are derived from the background region alone.
Now the background Stokes parameters can be used in Eq.~\ref{eq:pol_n0hat} (via the definition of $W_i$) to derive an equation involving the source Stokes parameters
similar to Eq.~\ref{eq:n0}
that can be solved iteratively for $\hat{N_0}$ for trial values of $\hat{\Pi}$ and $\hat{\varphi}$.

Finally, Eq.~\ref{eq:ps_ppi} is modified to be
\begin{equation}
    \frac{2}{\sigma_{\Pi}^2} = 2 \sum_{i=1}^{N} W_i^2 \mu_i^2 \cos^2 (2 \psi_i-2 \varphi)
   \label{eq:pol_ps_ppi}
\end{equation}
\noindent
to zeroth order in $\zeta^2 N_B / N$ and $\Pi_B^2 \equiv q_b^2 + u_b^2$.  To first order in these quantities,
\begin{eqnarray}
\begin{array}{l}
   \sigma_{\Pi}^2 \to \displaystyle \frac{1}{(N - \zeta N_B)^2} \left[ \frac{N^2}{\sum_i c_i^2} + \frac{\zeta^2 N_B^2}{\sum_j c_j^2} + \zeta^2 N_B {\rm\Pi}_B^2 \cos^2(2\varphi - 2\varphi_B)\right]  \\
   \ \approx \ \displaystyle \frac{1}{C_S^2} \left[ \frac{2(C_S+C_B)}{{\langle\mu_i^2\rangle}} + \frac{2 \zeta C_B}{{\langle\mu_j^2\rangle}} + \zeta C_B {\rm\Pi}_B^2 {\cos}^2(2\varphi - 2\varphi_B)\right] ,
\end{array}
   \label{eq:pol_sigmap}
\end{eqnarray}

\noindent
as $\Pi \longrightarrow 0$, with $\varphi_B = \frac{1}{2} \tan^{-1} (u_b/q_b)$.
The first term replicates Eq.~\ref{eq:sigmap}.
Because the extra terms are positive definite, they will  increase $\sigma_{\Pi}$, making the estimate of $\Pi$ more uncertain when there is polarized background, as expected.  The
magnitude of the increase in the uncertainty depends on the ratio of the expected polarized counts to the total counts in
the source region as well as the correlation between the source and background polarization phases.
}

\section{An Unbinned Model Test}
\label{sec:modeltest}
Consider a Kolmogorov test of conditional probabilities for a model where $q$ and $u$ depend on
$\xi$, representing time, spatial location, or energy.  For example, a model where the polarization fraction is constant with time while the EVPA rotates uniformly with rate $\omega$ could be specified as

\begin{eqnarray}
q(t) & ~=~ & \Pi \cos 2(\phi_0 + \omega t) \\
u(t) & ~=~ & \Pi \sin 2(\phi_0 + \omega t)
\end{eqnarray}

\noindent
where $\phi_0$ and $\omega$ are (fitted) parameters of the model to be tested, $\xi = t$, and each event has a specified value of $t$ given by $t_i$.  This model was applied to \ixpe\ data from Mk 421, finding rotation rates of $\omega = 80 \pm 9\deg$/d in one observation and $\omega = 91 \pm 8\deg$/d in another \citep{mk421rot}.

Generally, using the source region event density given by Eq.~\ref{eq:lambda_bg}, the conditional probability that $\psi \le \psi_i$ for event $i$ given that $\xi = \xi_i$ is

\begin{eqnarray}
C(\le \psi_i ~|~ q[\xi_i], ~u[\xi_i], ~\hat{N_0}, ~\hat{B}) & 
   ~=~ & \frac{\int_0^{\psi_i} \lambda(\psi; \xi_i) d\psi}{\int_0^{2\pi} \lambda(\psi; \xi_i) d\psi} \\
& ~=~ & \frac{\psi_i (1+\zeta N_B/\hat{N_0}) + 
    \mu_i( [q_i \sin 2\psi_i]/2 + u_i \sin^2 \psi_i ) }{2 \pi (1+\zeta N_B/\hat{N_0})}
\label{eq:cum_psi}
\end{eqnarray}

\noindent
where $q(\xi_i) \equiv q_i$ and $u(\xi_i) \equiv u_i$.  As ${\Pi} \longrightarrow 0$,
$C(\le \psi_i)$ approaches the uniform distribution, as expected.  Under the hypothesis that
the model is correct, though, we expect Eq.~\ref{eq:cum_psi} to give values that are uniformly
distributed between 0 and 1 even if $p$ is non-zero.  Thus, a Kolmogorov test of the cumulative
distribution of $C(\le \psi_i)$ values should provide a valid unbinned test of the event angles.

This test was implemented in Interactive Data Language (IDL) and applied to { several} different data sets from \ixpe.  In { each case}, events in the 2-8 keV band were used, the source region was 60$\arcsec$ in radius, and the background was taken from an annulus 200$\arcsec$ to 300$\arcsec$ from the point source.  The first source, Mk 501 (\ixpe\ data set 01004501), was found to be 10 $\pm$ 2\% polarized \citep{2022Natur.611..677L}.   For the null hypothesis that Mk 501 is unpolarized, the distribution of $C(\le \psi_i)$ deviated from the uniform distribution by 0.0085 with a set of 85,388 events in the source region; thus, the null hypothesis is rejected with a probability of less than $8\times 10^{-6}$.  { A likelihood ratio test rejects the null hypothesis with a probability of $7 \times 10^{-7}$ in this case, providing a somewhat better result for a simple test that the source is polarized.}  Under the hypothesis that the source is polarized, with parameters determined using the maximum likelihood method in \S~\ref{sec:modelbg}, then the deviation dropped to 0.00196, for a K-S probability of 0.90; thus, the constant polarized model { with fixed $\Pi$ and $\varphi$ } is acceptable, { a conclusion that was not available to \citet{2022Natur.611..677L}.}
{ Similarly, constant rotation models for the second and third \ixpe\ observations of Mk 421 (data sets 01003801 and 01003901, reported by \citet{mk421rot}) are accepted with probabilities of 0.97 and 0.78, respectively.}
{ Finally, the test was run on data from} Cen A (\ixpe\ data set 01004301), for which no polarization was detected; the upper limit to the polarization was 6.5\% at 99\% confidence \citep{Ehlert2022}.
For Cen A, the null hypothesis (that the source is unpolarized) is not rejected, giving a maximum deviation of 0.0039 with 28,078 events and a K-S probability of 0.79.
In summary, while an analysis may provide parameters of a polarization model, this test can be used on unbinned data to test the validity of the model, providing the user a diagnostic that could indicate whether the model is inadequate.

\section{Summary}

The unbinned likelihood method for X-ray polarimetry data analysis has been extended in several ways:

\begin{enumerate}

\item{Because many X-ray polarimeters must be rotated in order to be sensitive to arbitrary polarization position angles, an exposure weighting
approach was added.  A simple diagnostic term is developed that can inform the user when polarization measurements may be
deleteriously affected.}
\item{A way of accounting for background has been added to the basic formalism.  The background can be unpolarized but
it may be more common to have a polarized background, such as when observing a point source in a polarized nebula or near a brighter
polarized source.}
\item{An unbinned test using event phase angles was proposed that can be used to determine whether a
time- or energy-dependent model may be rejected.  The test was applied successfully to several \ixpe\ data sets.}
\end{enumerate}

\begin{acknowledgments} 

Funding for this work was provided in part by contract 80MSFC17C0012 from the MSFC to MIT in support of the \ixpe\ project.
This research used data products provided by the IXPE Team (MSFC, SSDC, INAF, and INFN) and distributed with additional software tools by the High-Energy Astrophysics Science Archive Research Center (HEASARC), at NASA Goddard Space Flight Center (GSFC).
Support for this work was provided in part by the National Aeronautics and Space Administration (NASA) through the Smithsonian Astrophysical Observatory (SAO) contract SV3-73016 to MIT for support of the Chandra X-Ray Center (CXC), which is operated by SAO for and on behalf of NASA under contract NAS8-03060.

\end{acknowledgments}

\facilities{\ixpe}
\software{ {Interactive Data Language (IDL)} }

\bibliography{main_revised}

\begin{center}
{\it Note added in proof}
\end{center}

Between acceptance of this paper and the review of the proofs, Eq.~\ref{eq:pol_sigmap} was independently derived using a Stokes
subtraction formalism and the equation was reformed in order to improve its interpretation.
While the first term dominates for weakly polarized background and for large source signals, the remaining terms deserve further comment.
The second term arises because of the variance of the background $Q$ and $U$, which is not considered in the case of unpolarized background.
The term does not depend on the actual value of the polarization of the background, which may be negligible.  Thus, {\em unless one has
independent evidence that the background polarization is zero}, Eq~\ref{eq:pol_sigmap} should be used instead of Eq.~\ref{eq:sigmap} or \ref{eq:mdp_bg0}.
The ratio of the second to the first term is approximately $\zeta C_B/ N$; if the source region has fixed size, this term can only be reduced by reducing $\zeta$ (i.e., increasing the size of the background region relative to the source region).
The third term does depend on the polarization of the background but is zero when the source EVPA is perpendicular to the background EVPA.  Thus, when the background is polarized, there is an asymmetry in the source polarization uncertainty contours in $q,u$ space.

\end{document}